\documentclass[conference,compsoc]{IEEEtran}

\usepackage{cite}
\usepackage{amsmath,amssymb,amsfonts}
\usepackage{algorithmic}
\usepackage{graphicx}
\usepackage{textcomp}
\usepackage{xcolor}

\usepackage{tikz}
\usepackage{caption}
\usepackage{subcaption}
\usepackage{url}
\usepackage{multirow}
\usepackage{listings}
\usepackage{adjustbox}
\usepackage{hyperref}
\usepackage{paralist}
\usepackage{tabularx}
\usepackage{etoolbox}

\usepackage{xcolor} % Add this line to include the xcolor package

\usepackage[utf8]{inputenc}
\usepackage[english]{babel}
\usepackage{hyperref}
\addto\extrasenglish{
    
}

\usepackage{ulem}
\usepackage{color,colortbl}
\usepackage{lineno}
\newcommand\redsout{\bgroup\markoverwith{\textcolor{red}{\rule[0.5ex]{2pt}{0.4pt}}}\ULon}

\newif\ifcomments
\commentstrue
\def\BibTeX{{\rm B\kern-.05em{\sc i\kern-.025em b}\kern-.08em
    T\kern-.1667em\lower.7ex\hbox{E}\kern-.125emX}}

\begin{document}

\title{On Feasibility of Server-side Backdoor Attacks on Split Learning}
\author{\IEEEauthorblockN{Behrad Tajalli}
\IEEEauthorblockA{\textit{ICIS, Radboud University} \\
% \textit{Radboud University}\\
% Nijmegen, Netherlands \\
hamidreza.tajalli@ru.nl}
\and
\IEEEauthorblockN{O\u{g}uzhan Ersoy}
\IEEEauthorblockA{\textit{ICIS, Radboud University} \\
% \textit{Radboud University}\\
% Nijmegen, Netherlands \\
oguzhan.ersoy@ru.nl}
\and
\IEEEauthorblockN{Stjepan Picek}
\IEEEauthorblockA{\textit{ICIS, Radboud University} \\
% \textit{Radboud University}\\
% Nijmegen, Netherlands \\
stjepan.picek@ru.nl}
}
\maketitle
\begin{abstract}
Split learning is a collaborative learning design that allows several participants (clients) to train a shared model while keeping their datasets private. 
In split learning, the network is split into two halves: clients have the initial part until the cut layer, and the remaining part of the network is on the server side.
In the training process, clients feed the data into the first part of the network and send the output (smashed data) to the server, which uses it as the input for the remaining part of the network.
Recent studies demonstrate that collaborative learning models, specifically federated learning, are vulnerable to security and privacy attacks such as model inference and backdoor attacks.
%Backdoor attacks are a group of poisoning attacks in which the attacker tries to control the model output by manipulating the model's training process. 
While there have been studies regarding inference attacks on split learning, it has not yet been tested for backdoor attacks.
This paper performs a novel backdoor attack on split learning and studies its effectiveness. Despite traditional backdoor attacks done on the client side, we inject the backdoor trigger from the server side. We provide two attack methods: one using a surrogate client and another using an autoencoder to poison the model via incoming smashed data and its outgoing gradient toward the innocent participants. %We did our experiments using three model architectures and three publicly available datasets in the image domain and ran a total of 761 experiments to evaluate our attack methods. 
The results show that despite using strong patterns and injection methods, split learning is highly robust and resistant to such poisoning attacks. While we get the attack success rate of 100\% as our best result for the MNIST dataset, in most of the other cases, our attack shows little success when increasing the cut layer.

\end{abstract}

\begin{IEEEkeywords}
Split Learning, Backdoor Attack, Cut layer
\end{IEEEkeywords}

%-------------------------------------------------------------------------------
\section{Introduction}
\label{sec:introduction}
Collaborative (also known as distributed) machine learning allows users to train a shared model together without putting their private dataset at risk and sharing it directly with other parties while benefiting from a more extensive volume of data~\cite{DBLP:journals/csur/VerbraekenWKKVR20, DBLP:conf/ccs/ShokriS15}. Several solutions have been proposed for collaborative learning, with two well-known ones being \textit{Federated learning} (FL)~\cite{DBLP:journals/corr/KonecnyMYRSB16} and \textit{Split learning} (SL)~\cite{DBLP:journals/corr/abs-1812-00564}. 
% FL has been proposed in two different designs according to their usage for training parties: Vertical Federated Learning (VFL) and Horizontal Federated Learning (HFL) \cite{DBLP:journals/kbs/ZhangXBYLG21}. HFL, which is the most common design, allows the parties to train their own model with their private dataset and, when finished, aggregate their trained models through a trusted central server.
The architecture in both FL and SL comprises a central remote server and several participants (called clients). The clients do the first round of computations and send their computed results to the server afterward. The server is usually responsible for aggregation (FL) or performing the rest of the computation task (SL). 

In SL, the network is usually split into two subnetworks between the client and server. The server has the upper part of the network (which usually has more layers and bears most of the computation cost), and the client has the lower part. The client processes the input until its cut layer and then sends intermediate computation (called \textit{smashed data}) to the server and receives the gradients in return (Figure~\ref{fig:SL_training} in \autoref{sec:appendix2}). Similar to FL, SL provides the same advantage of users' raw data not being shared directly with each other. Fewer computation costs on the client side and less communication overhead and power consumption are other main advantages of using SL rather than FL~\cite{DBLP:journals/corr/abs-1812-03288, DBLP:conf/srds/GaoKAKTKCKN20}. Since SL has been recently proposed, few studies have been conducted on its security and privacy. Still, existing works have revealed its vulnerabilities to inference attacks~\cite{DBLP:conf/ccs/PasquiniAB21, DBLP:conf/acsac/HeZL19}. 
%\textcolor{red}{A recent research~\cite{khan2022security} has performed an empirical analysis of SplitFed~\cite{thapa2022splitfed} robustness to model poisoning attacks.}\todo{and so what? still not clear how connected or what they found???}

Backdoor attacks are a group of poisoning attacks in which the attacker aims at making the model learn a specific type of pattern in input, so whenever this pattern is added to input during test time, the model outputs the desired target class and otherwise functions as expected. This is typically done by poisoning part of the training dataset with patched inputs and training the model on the whole dataset (including clean and poisoned samples). 

This work investigates whether SL is vulnerable to backdoor attacks. Here, we can consider two scenarios. The first scenario is a conventional backdoor attack in which the attack is performed from the client side.
The malicious attacker can access one or multiple clients and manipulate their dataset and/or model. The main challenge in this attack is dealing with the cancellation effect, which stems from the round-robin training method. The second challenge is to bypass detection and filtering mechanisms deployed on the server side. With that said, a backdoor attack from the client side seems feasible. This type of attack is limited to the case where the attacker is one of the clients (or has access to its data). 
Yet, there is another scenario in which all clients can be honest, and the attacker can be on the server side. %For instance, imagine that most institutes and companies know each other. There should be a highly trusted environment between them that can be used to train an SL model on their sensitive data.
Furthermore, the clients' databases are well protected physically and logically. This raises a question: \textit{Can an attacker perform a backdoor attack on the SL design from the server side?} Our research aims to provide that answer and examine SL vulnerability to backdoor attacks conducted through the server side.

The main challenge is that, as attackers, we have access to neither input data to manipulate nor a lower subnetwork to modify directly. The only means by which we could affect the subnetwork are the gradients that come back from the server. To this end, we tried two different backdoor injecting methods to observe if there was any success during test time. On our first try, we trained a surrogate client model simultaneously while the innocent clients were being trained. The input for this surrogate client is a poisoned dataset, and its output would be sent to the server. The loss function is the average of the innocent and surrogate client computations, which will be propagated back to both clients in each iteration. 
In the second, more advanced method, we use a trained autoencoder as a transformer to inject backdoored smashed data into the server subnetwork and affect the client subnetwork consequently. Our autoencoder is trained with a dataset consisting of couples of $(clean\ smashed\ data, backdoored\ smashed\ data)$ as its (input, label) pair. Thereafter, we install the autoencoder between the innocent client output and server input while training, and the backpropagation data would go directly from server to client. With these two approaches, we aim to see if the gradient coming back from the server could affect the innocent client parameters during its gradient descent update step. %In the test time, the backdoor pattern would achieve an acceptable ASR. 
Our results indicate both methods were unsuccessful in injecting the backdoor and poisoning the split learning model. Yet, in most cases, the clean accuracy of the model remains high even after backdoor injection. The main contributions of this paper are summarized as follows: 
\begin{compactitem}
    \item To the best of our knowledge, this is the first study that investigates the SL backdoor injection attack from the server side and defines a new threat model and scenario in which a backdoor trigger is injected without access to the model's input and only tampering with the midway passing (smashed) data.
    \item We perform our attack with the best-case scenario in which the poisoning ratio and trigger sizes are large enough (to achieve 100\% success in any other backdoor studies) and display the resistance and stability of SL design against backdoor attacks while maintaining the original task accuracy high.
    \item We perform two types of attacks to conclude the robustness of SL against server attempts: direct attack via injecting the poison data from a surrogate client and indirect attack by installing an autoencoder between the server and innocent clients.
\end{compactitem}

%-------------------------------------------------------------------------------
\section{Background}
\label{sec:background}
%-------------------------------------------------------------------------------

% \subsection{Collaborative Learning}
% In addition to preserving users' data privacy, Collaborative Learning provides a training platform for cases with limited data for each user. To this end, multiple clients will provide their collected dataset to a shared learning platform and could benefit from training a more complex model on a larger dataset. Each of the clients has its private dataset. After the training is completed, all have access to a shared model. For example, suppose $n$ hospitals collected a dataset of tumor pictures and their diagnosis labels from their own private patients and strictly do not want to share their patients' data with anybody. We consider a set of remote clients $C_r=\{c_1, ... , c_n\}$ and each client $c_i$ has its own private dataset $X\mbox{-}priv_i $. This would contribute to training a shared model $M$ that will learn the distribution of data samples among all $X\mbox{-}priv_i$ and is expected to classify the tumor pictures accurately in their correct category. FL is the most famous collaborative learning solution for which Several implementations have been proposed. \autoref{sec:hfl_training_prcss} in \autoref{sec:appendix2} explains in more detail how a typical HFL training routine goes on.  

\subsection{Split Learning}
In split learning, the neural network is partitioned into sequential stacks of layers (usually 2 or 3 sub-networks). There is a central server, and multiple remote clients participate in the learning process. In the start-up phase, remote clients will reach a consensus about the learning task, the neural network's architecture, how to partition the neural network, and other hyperparameters. The server has no role in this decision-making process. Then, they will send the setup information to the server. We consider a set of remote clients $C_r=\{c_1, ... , c_n\}$ and each client $c_i$ has its own private dataset $X\mbox{-}priv_i $. This would contribute to training a shared model $M$ that will learn the distribution of data samples among all $X\mbox{-}priv_i$. 

The Vanilla SL training protocol runs as follows:
The local remote client $c_i$ downloads the global updated $M_c$ (with $M_c$ being the client part of the model and $M_s$ the server part) and computes the output of its sub-network (performs forward propagation up to its final layer called the cut layer). Here, the cut layer output is called smashed data ($smsh_i = M_c(X\mbox{-}priv_i)$). Then it sends the smashed data ($smsh_i$) to the server.
The server propagates forward to the end of its network layer ($M_s(smsh_i)$).
Now, the loss function shall be calculated concerning ground-truth labels $Y_i$ (i.e., $\mathcal{L}_M=L(M_s(smsh_i), Y_i$). The server has access to the ground truth labels. Thus, it calculates the loss and backpropagates down to its initial layer. Then, it will send the gradients back to client $c_i$.  
Finally, the client $c_i$ performs the rest of the backpropagation and then uploads the updated $M_c$ for the next client $c_{i+1}$ to use.
Each training step is completed when one full round-robin circle is achieved among all clients. The client and server models are stored during and after training, usually in a trusted third party. Thus, the global model could always be downloaded from or uploaded to this third server.

SL has been shown to have similar privacy features as those in FL~\cite{DBLP:journals/corr/abs-1812-00564}. Less computational overhead, communication costs, and power consumption are other benefits of using SL over horizontal FL~\cite{DBLP:journals/corr/abs-1812-03288, DBLP:conf/srds/GaoKAKTKCKN20}. Yet, SL is not the perfect collaborative solution free from any security vulnerabilities. As it is quite a new concept, extensive research has not yet been done studying its security and privacy issues. %Few studies have been conducted so far, showing that SL is highly susceptible to inference attacks~\cite{DBLP:conf/ccs/PasquiniAB21, DBLP:conf/acsac/HeZL19}.

\subsection{Backdoor Attacks}
Backdoor attacks are a type of poisoning attack whose purpose is to implant a trigger in the model during the training phase. In data poisoning, the backdoors are injected into a model by poisoning the training data with a trigger pattern. Once the training is complete, the poisoned model is expected to function normally on clean inputs. Still, whenever the poisoned input feeds the model, the trigger is activated within the model, causing it to predict the poisoned input as a target class~\cite{DBLP:journals/corr/abs-2007-08745}. This is mainly because some neurons within the model have learned the trigger pattern, and their activation values rise drastically when fed by poisoned input, causing the model's last layer to have the highest confidence in the target class rather than the valid one~\cite{DBLP:journals/corr/abs-2107-07240}. 
% The malicious attacker could determine the target class even without manipulating the labels \cite{DBLP:conf/nips/ShafahiHNSSDG18}. The backdoor trigger patterns could be anything interpretable by the model (e.g., random or calculated pixel patterns in computer vision \cite{DBLP:journals/corr/abs-1712-05526}, some particular phrase in text classification \cite{DBLP:conf/ndss/LiuMALZW018}, a tone or frequency in speech recognition \cite{DBLP:conf/wisec/KoffasXCP22}, and so on).

Several research works have been done to evaluate the vulnerability of collaborative learning to client-side backdoor attacks, specifically those done on FL systems~\cite{DBLP:conf/aistats/BagdasaryanVHES20, DBLP:conf/nips/WangSRVASLP20, DBLP:journals/corr/abs-1911-07963, DBLP:conf/iclr/XieHCL20}. The core idea behind most of these attacks on FL is that malicious clients train their models with backdoored data and inject the backdoor into the global model by providing poisoned parameter updates.
Recently, a poisoning attack against SplitFed architecture~\cite{thapa2022splitfed} is presented in~\cite{khan2022security}.
SplitFed is a hybrid architecture combining SL and FL where clients interact with the SL server separately and then update their model using the FedAvg server. 
As mentioned in the paper~\cite{khan2022security}, compared to FL, it is challenging to poison the SplitFed structure because the attacker does not have access to the upper part of the network.
The authors present a client-side optimization-based poisoning attack and leave the server-side poisoning attack as future work.

% The core idea behind the design of most attacks against FL is as follows: First, the attacker tries to poison one or multiple models on the client side (in their threat model, the assumption is that the attacker can manipulate one or multiple clients and has access to their models and also datasets). Then on each model update phase, they try to replace the global model with the malicious one. Since in each aggregation, the effect of the malicious model would be canceled out by other innocent updates; the attacker would multiply the parameters he is sending by some boost factor $\gamma$ to ensure that the backdoored model will survive the aggregation process.

%-------------------------------------------------------------------------------
\section{Backdoor Attacks on Split Learning}
\label{sec:backdoor_attack}
\subsection{Threat Model}
%We assume that there is only one participant which is considered an attacker. 
The attack is performed in the Vanilla SL model with $K\geq 1$ clients and one server, where the server has the labels.
We assume the attacker controls the server side and all clients are honest.
The clients have a homogeneous dataset from the same distribution. % (e.g., all hospitals have a collection of different 2D tumor pictures).

$\blacktriangleright$ \textbf{Attacker's goal:} The attacker aims to successfully backdoor the whole structure of the neural network (both client and server parameters). A backdoored model should satisfy (i) high attack accuracy, i.e., when facing the poisoned input, it should classify the input as a backdoored target class,  and (ii) high clean data accuracy, i.e., when facing the typical clean input, it should predict the ground truth labels with high probability. %with that said, as an attacker, we need both a high attack success rate and clean data accuracy during test time. 

$\blacktriangleright$ \textbf{Attacker's knowledge:} We assume that the attacker knows the clients' subnetwork architecture and the dataset distribution of clients. This means the attacker has its own local dataset, which is the same size and from the same distribution as other participants on the client side. Furthermore, the attacker knows the label of the incoming smashed vector since it has access to all labels provided by all participants.

$\blacktriangleright$ \textbf{Attacker's capability:} %No malicious entity could have access to the clients' clean inputs, models, or any of their hyperparameters. 
    The attacker has access to the intermediate computations between client and server (here, smashed data and backward gradients) and ground-truth labels sent to the server in each iteration (for loss computation).
    Additionally, the attacker has full access and control over the upper subnetwork parameters.
    However, since the clients are honest, the attacker cannot see or tamper with their datasets, models, or hyperparameters. 
    The attacker can only modify the lower subnetwork parameters through the backward gradient passed through the client after loss computation during each iteration.

\subsection{Proposed Attacks}
%We now introduce our server-side attacks, where we propose two different methods to inject the backdoor in the split learning model. 
%First, we deploy a surrogate client model. Second, we train an autoencoder to poison the model with our backdoor trigger.

\subsubsection{Surrogate Client}
\begin{figure}
    \centering
    \includegraphics[scale=0.06]{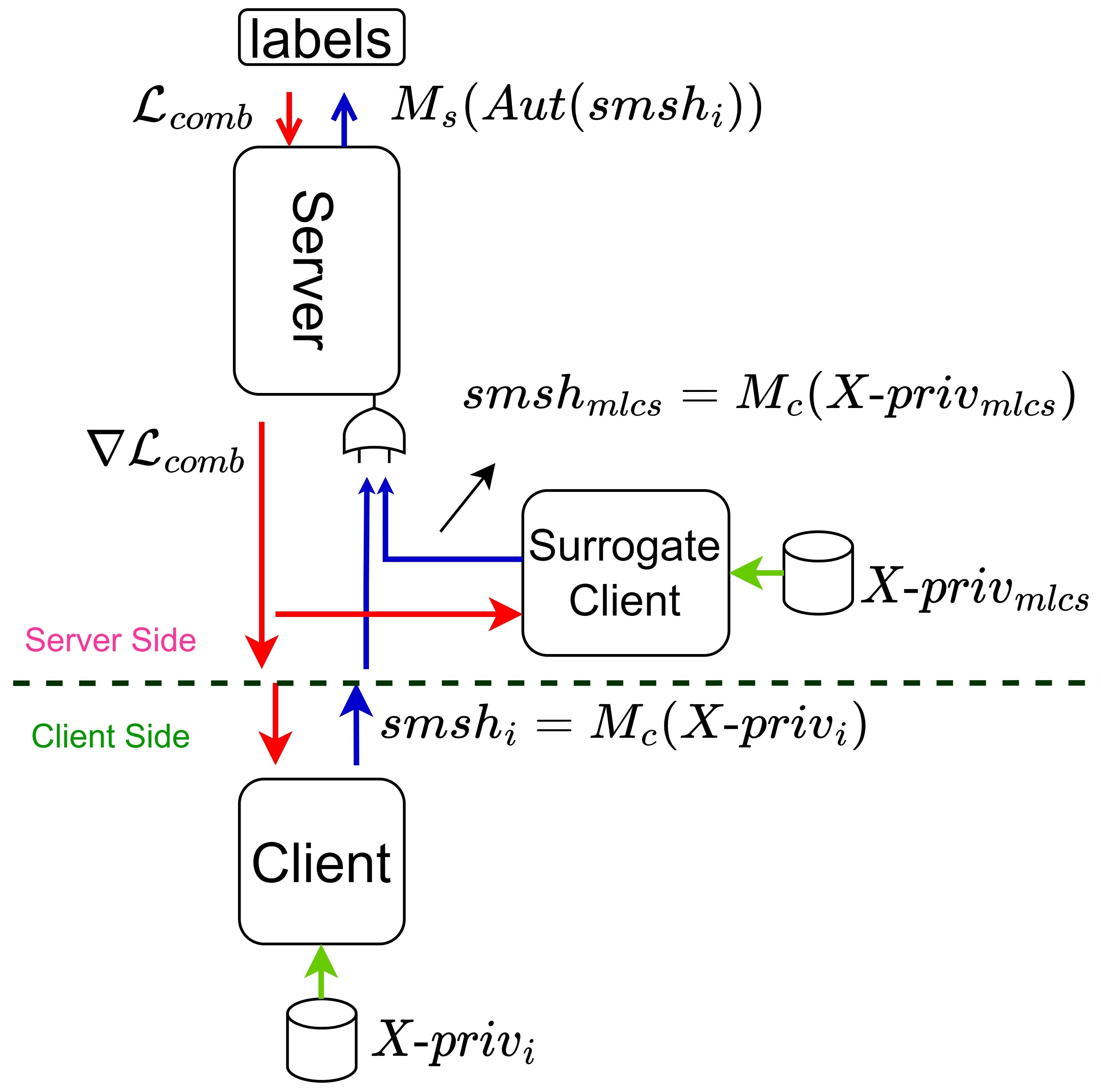}
    \caption{First attack scenario by setting up a surrogate client on the server side. }
    \label{fig:surrogate_client_attack}
\end{figure}

In this scenario, we aim to inject the backdoor trigger utilizing a surrogate client on the server side, which the (honest) clients do not know (see \autoref{fig:surrogate_client_attack}). The surrogate client has the same model architecture as the other clients, but the initialization of parameters would be different due to no access to the clients' models. When the training has begun, in each forward pass, the surrogate client computes smash data that comes from a local poisoned dataset ($smsh_{mlcs} = M_c(X\mbox{-}priv_{mlcs})$). Afterward, the server computes the rest of the layers and calculates the loss value ($\mathcal{L}_{mlcs}$). The server had already computed the forward pass for an innocent client and also had calculated its loss ($\mathcal{L}_{incnt}$). Now, the server averages and backpropagates the combined loss of these two separate losses and sends it back to both innocent and surrogate clients. The combined loss is calculated through some average function. Here, we use simple averaging by using a weighting parameter $\alpha$:
\begin{equation}
        \mathcal{L}_{comb} = \alpha\mathcal{L}_{incnt} + (1-\alpha)\mathcal{L}_{mlcs}.
\end{equation}

\subsubsection{Injector Autoencoder}
In our second attempt, we tried to inject the backdoor by training an autoencoder and putting it in between the innocent client and server while training so that the autoencoder could transfer the innocent smash data to the poisoned one. For clarity, we explain the attack in several phases:

\begin{figure*}[ht!]
     \centering
     \begin{subfigure}[b]{0.3\linewidth}
         \centering
         \includegraphics[scale=0.07]{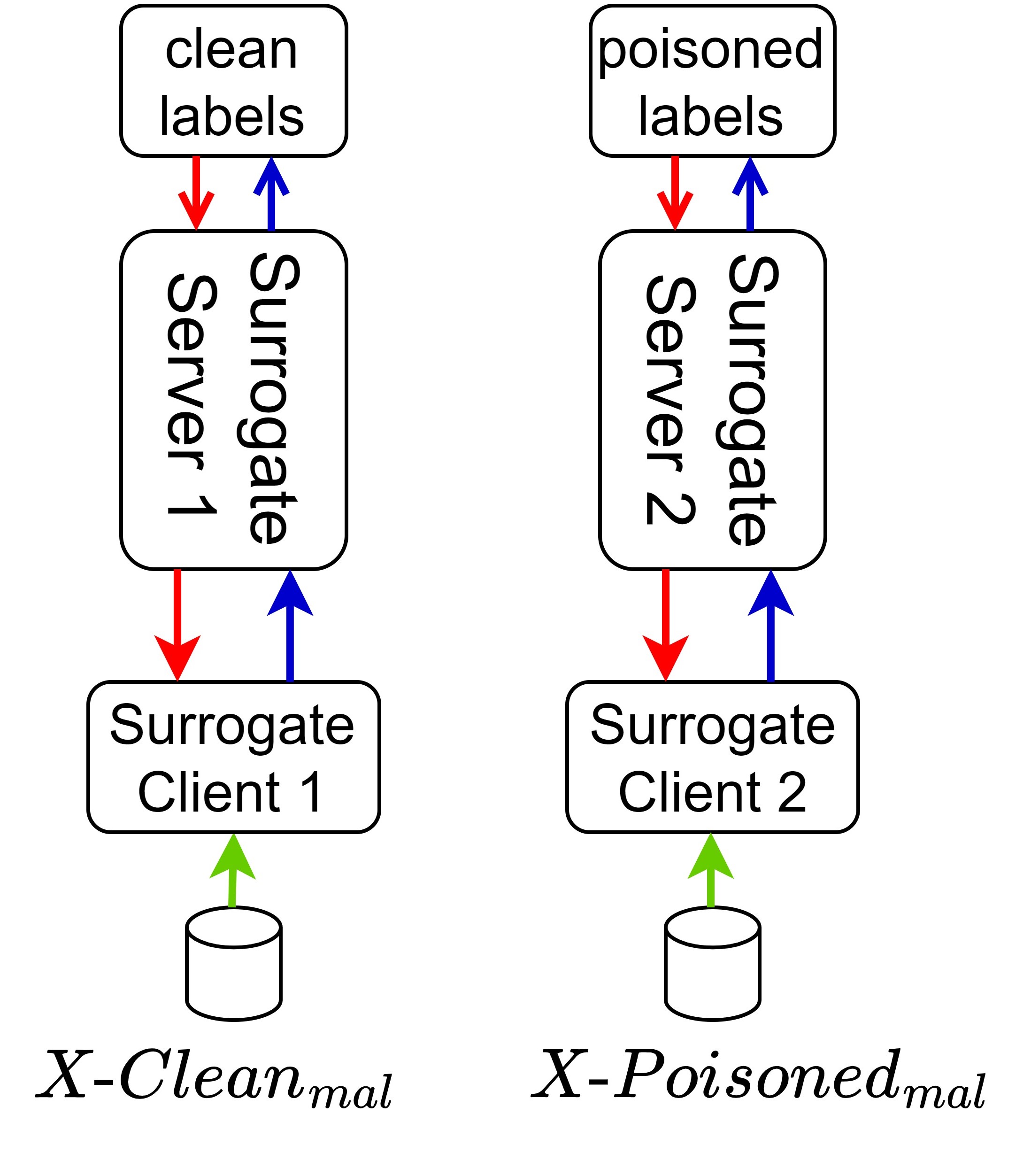}
         \caption{Phase 1: Training two surrogate models with the poisoned and clean dataset.}
         \label{fig:phase1}
     \end{subfigure}
     \hfill
     \begin{subfigure}[b]{0.3\linewidth}
         \centering
         \includegraphics[scale=0.07]{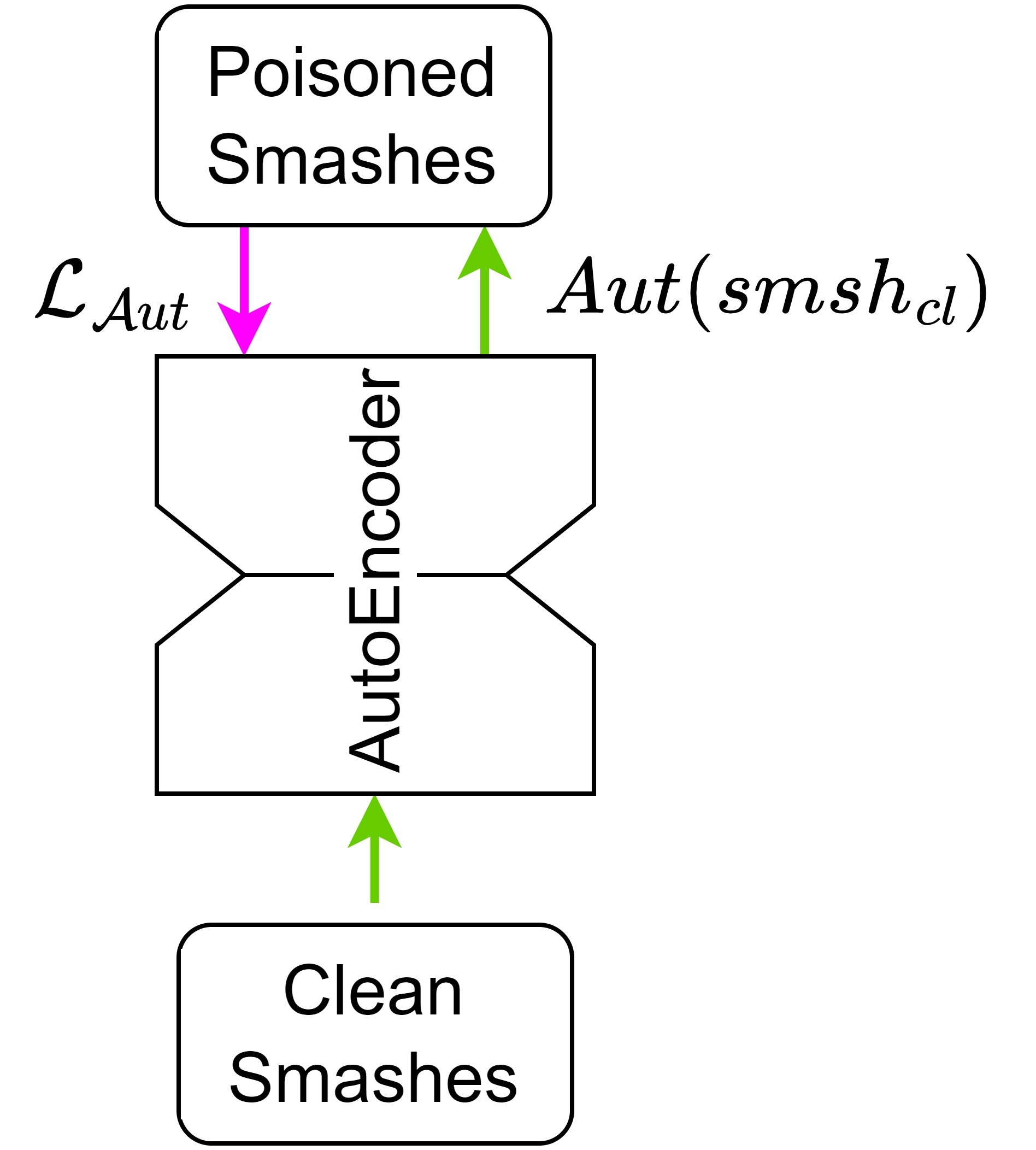}
         \caption{Phase 3: Training an Autoencoder by collected ($(clean_{smsh},  poisoned_{smsh})$ during previous phase.}
         \label{fig:phase3}
     \end{subfigure}
     \hfill
     \begin{subfigure}[b]{0.3\linewidth}
         \centering
         \includegraphics[scale=0.065]{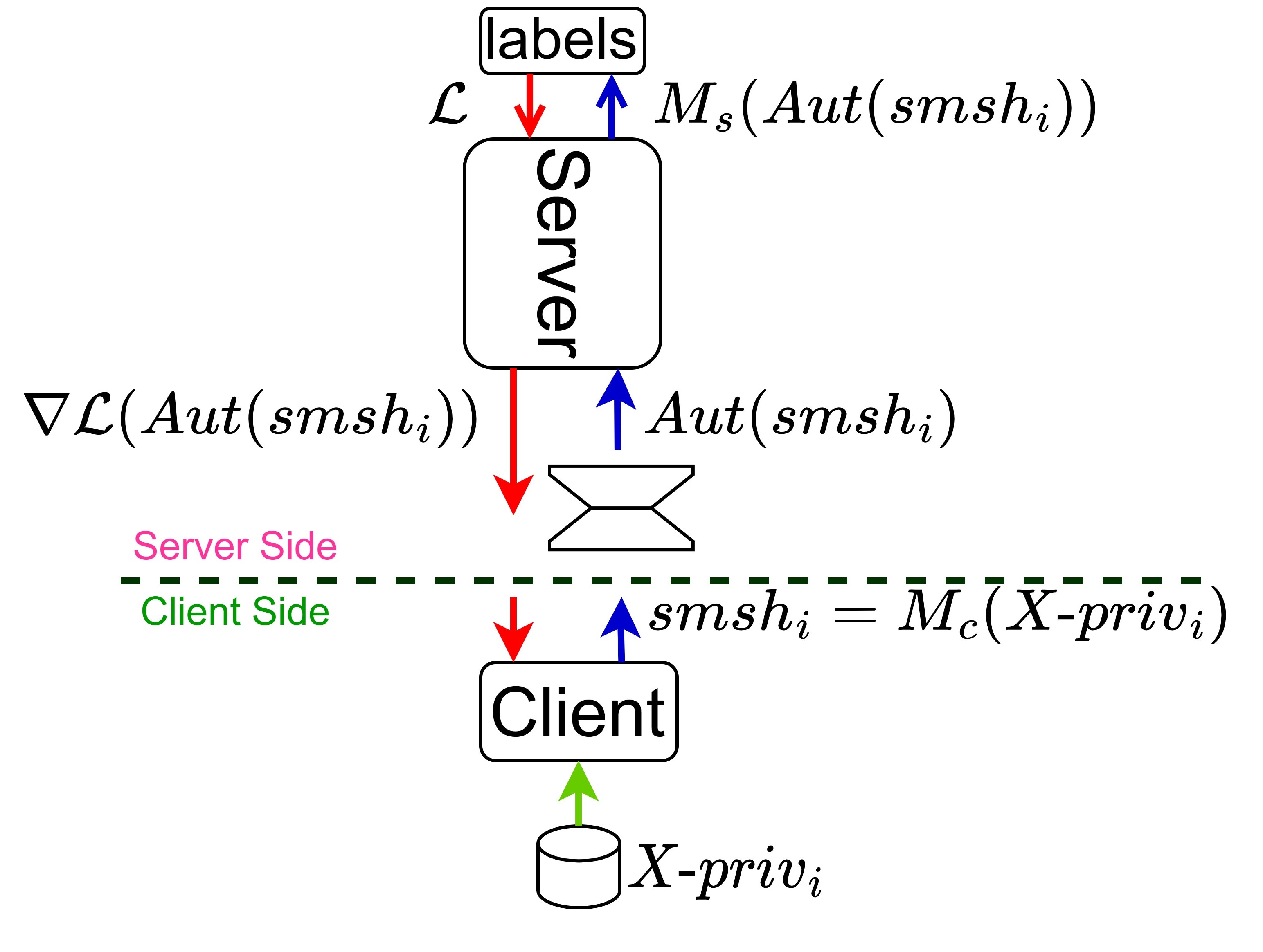}
         \caption{Phase 4: Inserting an Autoencoder between client and server during normal training.}
         \label{fig:phase4}
     \end{subfigure}
        \caption{Three (out of four) phases of our second attack using autoencoder.}
        \label{fig:autoencoder_attack}
\end{figure*}

     \textbf{1) Training surrogate networks:} In the initial phase, the attacker replicates two learning models precisely the same as the designated SL system but only with one client (the attacker can replicate the client subnetwork as well since he knows the architecture of the client model). Then, he trains each of these two separately. Using the attacker's own dataset, the first network is trained using the clean dataset (clean input and ground-truth label). Then, for the second network, the attacker poisons the dataset with his crafted backdoor and trains it similarly (Figure~\ref{fig:phase1}). 
    
    \textbf{2) Collecting a smashed dataset:} In the second phase, we test both networks as follows: the attacker feeds the clean input and its equivalent poisoned one to clean-trained and backdoor-trained networks, respectively. For those inputs whose backdoors are successful, we collect the clean smashes and their corresponding backdoored smashes as a pair of (Input, Label) samples. 
    
    \textbf{3) Training an Autoencoder:} In the third phase, we train an autoencoder with the data collected in the previous phase. The autoencoder is expected to generate backdoored smashed data given its corresponding clean one (Figure~\ref{fig:phase3}).

    \textbf{4) Deployment:} In the final phase (see Figure~\ref{fig:phase4}), we incorporate the trained autoencoder as a transformer at the cut layer. Note that the client is not aware of the autoencoder. During the regular training, the client computes forward pass up to its cut layer. Then sends the smashed data to the server. On the server side, the smashed data received from the client is fed to the autoencoder. The autoencoder generates the corresponding backdoored smashed data, which is now fed to the server subnetwork. The server backpropagation computations are sent directly back to the client. This way, we aim to inject the backdoor into both the server and the client subnetworks (through gradients).

%-------------------------------------------------------------------------------
\section{Experimental Setup}
\label{sec:experimental-setups}
%-------------------------------------------------------------------------------

Experiments were implemented and run using PyTorch v1.12 on a GPU cluster with CentOS Linux with NVIDIA GPUs (Tesla P100, GeForce GTX 1080 Ti, GeForce RTX 2080 Ti, and Tesla v100). More details on experimental settings can be found in \autoref{sec:appendix1}.

We perform our experiments using three publicly available datasets (MNIST, FMNIST, and CIFAR10) and three model architectures: LeNet~\cite{lecun1998gradient}, StripNet~\cite{strip}, and ResNet9~\cite{he2016deep}.
We consider two metrics to evaluate our experiments: Attack Success Rate (ASR) which indicates how much the backdoor is succeeding on the poisoned dataset, and Clean Data Accuracy (CDA) which measures the accuracy of the poisoned model on clean input.
For all experiments, we use a square shape $8\times8$ pixel trigger pattern with random pixel values injected in the upper middle part of the images.
We use the following hyperparameters:
\begin{compactitem}
    \item \textbf{Surrogate client attack:} For reproducibility purposes, in all experiments, we fix the seed to 47, and 0 is the target class label for all attacks. We fix the batch size to 128 and shuffled the data loader at the beginning of each epoch. Cross Entropy loss was our criterion for loss calculation. ADAM was chosen as the optimizer with an initial learning rate of 1e-3 and weight decay set to 1e-4. We use our custom-defined learning rate scheduler to decrease the rate by a factor of 1/10 depending on epoch number and model architecture. We use early stopping with patience of 70 for CIFAR10 and 90 otherwise. The number of epochs was set to 140 in the case of CIFAR10 and 90 otherwise.
    \item \textbf{Autoencoder injection attack:} We use the same setup with the surrogate client attack except for the followings: for training autoencoder, we use mean square error loss as our criterion and REDUCELRONPLATEAU~\cite{reducelronplateau_2022} as our learning rate scheduler with the patience of 20, factor of 0.2, and minimum learning rate of 5e-5. 
For the first training phase, the number of epochs is set to 100 for CIFAR10 and 70 otherwise. For the second phase (training autoencoder), the number of epochs is set to 180 for CIFAR10 and 100 otherwise.
\end{compactitem}

%-------------------------------------------------------------------------------
\section{Evaluation and Results}
\label{sec:evl_results}
%-------------------------------------------------------------------------------
\autoref{fig:baseline_clean_acc} in \autoref{sec:appendix3} shows the results of the models trained on clean data (test accuracy) that we use as a baseline accuracy (BA) to compare with the CDA of the models trained with poisoned samples.

\subsection{Results for Surrogate Client Attack}
\begin{figure}
    \centering
    \includegraphics[width=\linewidth]{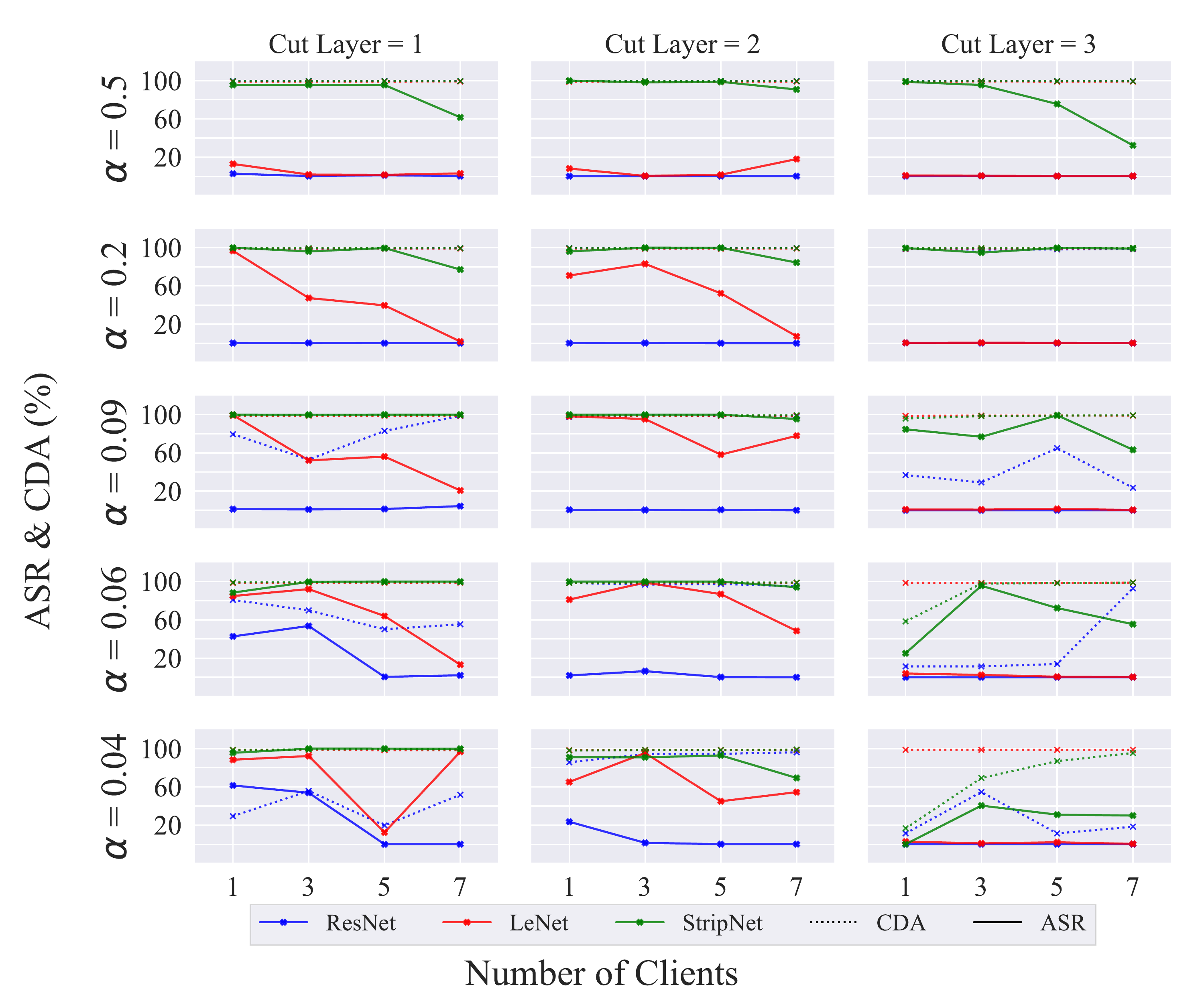}
    \caption{ASR and CDA for MNIST (surrogate client).}
    \label{fig:asr_mnist}
\end{figure}

\begin{figure}
    \centering
    \includegraphics[width=\linewidth]{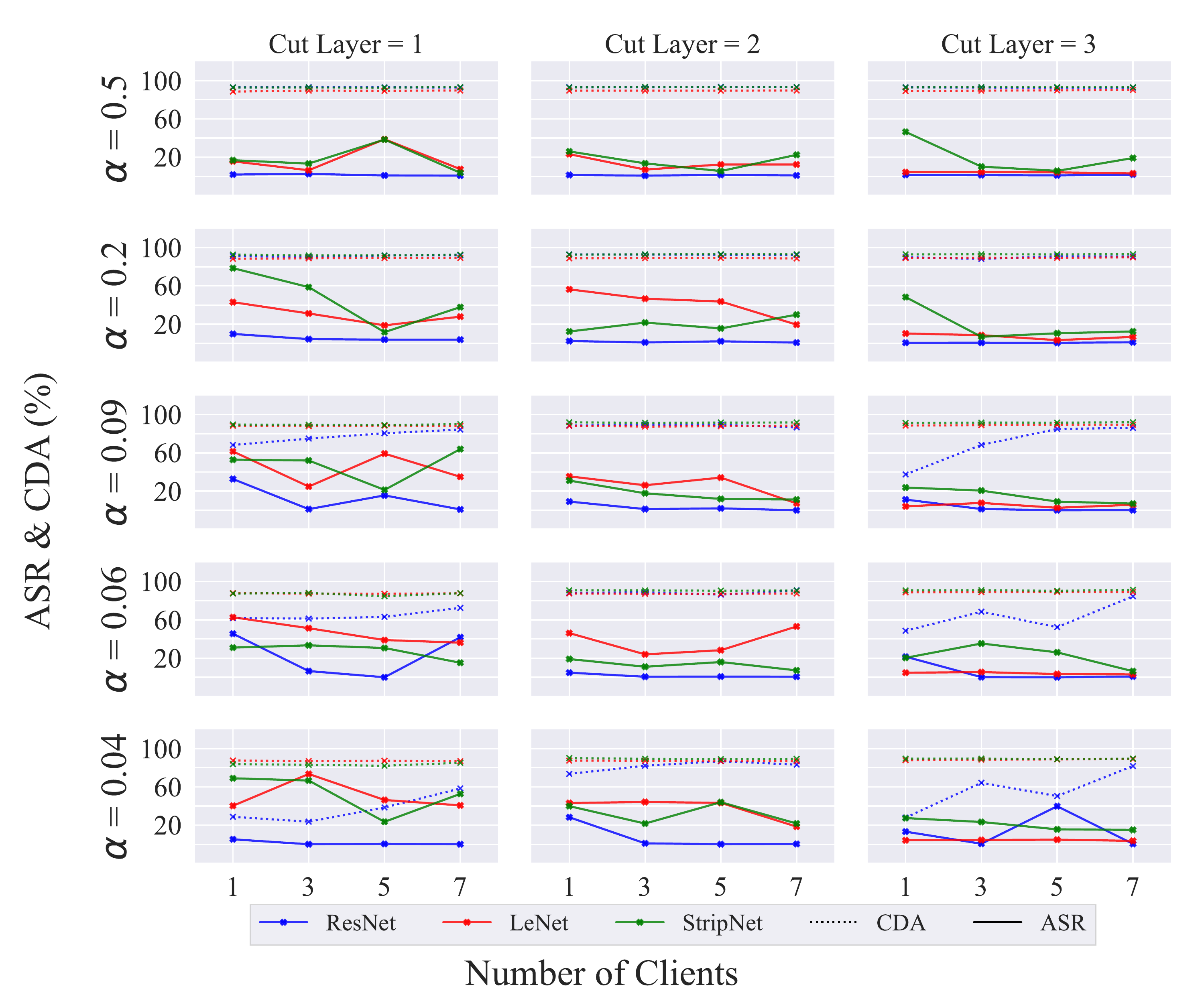}
    \caption{ASR and CDA for FMNIST (surrogate client).}
    \label{fig:asr_fmnist}
\end{figure}

\begin{figure}
    \centering
    \includegraphics[width=\linewidth]{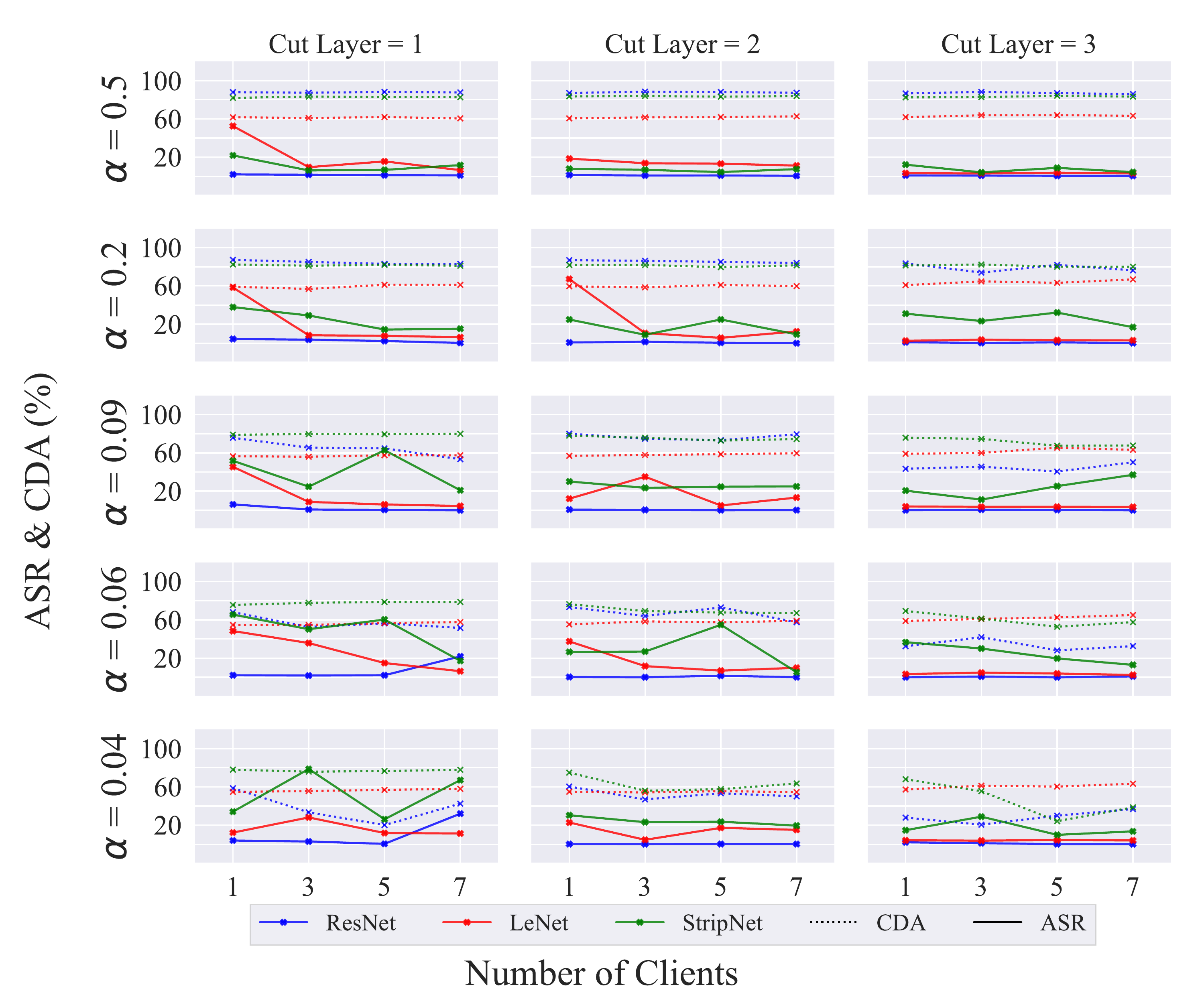}
    \caption{ASR and CDA for CIFAR10 (surrogate client).}
    \label{fig:asr_cifar10}
\end{figure}

\autoref{fig:asr_mnist}, \autoref{fig:asr_fmnist}, and \autoref{fig:asr_cifar10} present our experimental results for the surrogate client attack. Each figure contains $5\times3$ subfigures representing the obtained results (ASR and CDA based on the number of clients in increasing order) for a particular cut layer and $\alpha$ value. %Here, we discuss the results for each model architecture.


\noindent\textbf{ResNet:} For the MNIST dataset, ResNet achieves high CDA for cut layer 2, but for cut layers 1 and 3, while decreasing $\alpha$, CDA also drops significantly. These results hold while ASR is almost close to zero for all attempts (except 2 cases in cut layer 1 and $\alpha$ is 0.04 and 0.06, where ASR reaches up to 60\% in the best case). 
We see a similar trend for FMNIST. For cut layer 2, CDA and BA are the same, but for cut layers 1 and 3, as $\alpha$ decreases, CDA also drops noticeably. It seems that increasing the number of clients has a positive effect on CDA. We observe that ASR fails to reach even a random guess threshold (except for a few cases in cut layer 1, where it reaches an ASR between 40 to 50). 
CIFAR10 results also indicate a failure for our attack considering both ASR and CDA evaluation metrics.

\noindent\textbf{LeNet:} For the MNIST dataset, LeNet achieves high ASR (same as BA) in almost all cases. It can be observed that the results for cut layers 1 and 2 demonstrate the same trend, while for cut layer 3, it shows a complete failure of the attack. Increasing the number of clients negatively affects ASR (cut layers 1 and 2) in most cases. Still, in some cases switching between 5 and 7 shows a positive effect (e.g., cut layer 1, $\alpha$=0.04). With a decrease in alpha, there is a jump in ASR between $\alpha$=0.5 and $\alpha$=0.2 and, afterward, a mild improvement where it can reach more than 90\% in some cases when $\alpha$=0.04. This observation shows that the $\alpha$ parameter we used can effectively attack the LeNet model on MNIST.
For the FMNIST dataset, the result of CDA is as high as BA in almost all cases. For ASR, however, there are no promising results in most cases (for cut layer 3, the results are close to 0 in most cases). However, we could see some successful attempts in lower cut layers and fewer alpha values (73.5\% for $\alpha$=0.04 and cut layer1).
For CIFAR10, again, there is a successful trend in CDA results. We can observe that CDA stays around 60\% in almost all cases; in some cases, even better than BA (e.g., with cut layer 3 and 3 clients, it reaches up to 64.68\% for $\alpha=0.2$). The attack success rate is again not promising overall, but some very high ASRs when using just one client (mainly in 1 cut layer mode) are even higher than their corresponding BAs.

\noindent\textbf{StripNet:} For the MNIST dataset, StripNet could achieve high CDA (same as BA) for most cases except the ones in cut layer 3 (for $\alpha$=0.04, we observe a clear trend in improving CDA while increasing the number of clients). ASR in cut layers 1 and 2 is very high in almost all cases, but in cut layer 3, it is lower. For almost all plots, there is a drop in ASR when switching from 5 to 7 clients. The critical phenomenon in cut layer 3 is the drop in ASR when decreasing $\alpha$, which means the attack is not effective for these $\alpha$ values.
For FMNIST, CDA is high and very close to BA in all cases. Again, we cannot observe a successful attack in most cases except those with just one client (mostly in cut layer 1 with the highest ASR of 78.6\%). Other than in a few cases, a decrease in $\alpha$ has no direct relationship with increasing ASR.
In CIFAR10, we can observe that CDA is as high as BA, and by decreasing $\alpha$, it drops noticeably in most cases (specifically in cut layer 3). The parameter alpha seems to be effective in this case. By reducing $\alpha$, we could observe a trend in the improvement of ASR (particularly from 0.5 to 0.2). Nevertheless, results are still not promising and have fluctuating behavior. The best results are in cut layer 1 (78.7\% as the best ASR).

To conclude, except for MNIST, our attempts to inject a backdoor fail to achieve successful ASR. Nonetheless, CDA values remain in a reasonable range. 
We believe that the attack is reasonably successful for MNIST because of the simplicity of the dataset. Our reasoning is supported with further discussion and experimental results in Appendix~\ref{sec:appendix4}.

\subsection{Results for Autoencoder Injection Attack}
\begin{figure}
    \centering
    \includegraphics[width=\linewidth]{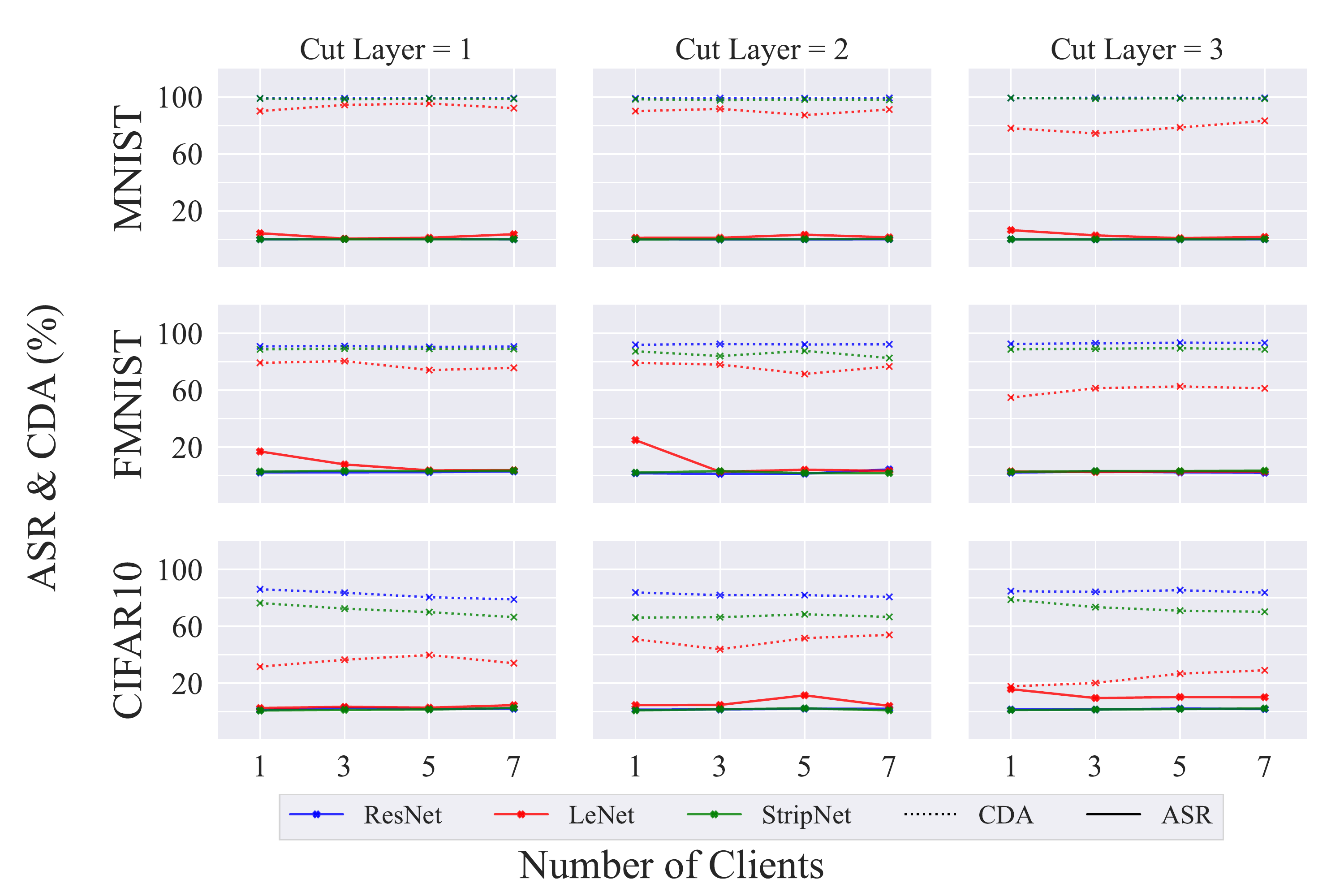}
    \caption{ASR and CDA for autoencoder scenario.}
    \label{fig:asr_autoencoder}
\end{figure}

\autoref{fig:asr_autoencoder} demonstrates the results of our second attempt to inject backdoors using an autoencoder. CDA values are high and almost identical to their corresponding BAs in all cases. However, ASRs are not promising and close to zero for most points (except for some values for LeNet, which is close to the random guess). The fact that ASR values are mostly below random guess demonstrates that the autoencoder output contains some information that makes the model learn another distribution regarding the  poisoned inputs, which causes unlearning effect for the distribution of poisoned data and places them in a different manifold. In future work, it would be interesting to determine the autoencoder's underlying function and the distribution to which it sends the given input.
%\onote{the reader want to know why it does not work, explaining the figure is repeating the results, but not explaining 'why' part}

%-------------------------------------------------------------------------------
\section{Discussion on the Attack Feasibility}
\label{sec:discussion}
%-------------------------------------------------------------------------------
%

This section provides a discussion on the difficulty and feasibility of injecting a backdoor from the server side.
Specifically, we discuss the challenge of influencing the client-side network via backpropagation.
Our explanation is generic and for simple datasets like MNIST, it may not be necessary for the success of the attack.

For simplicity, let us suppose that each layer in the network structure consists of just one neuron. This explanation is also valid for convolutional neural networks because the kernels have almost the same computation process as neurons. We assume we split the network from its $i_{th}$ layer to server and client subnetworks. With that, the output of the client ($a^{(i)}$) is as follows:
\begin{equation}
\begin{split}
    a^{(i)} = \sigma(z^{(i)}), \quad
    z^{(i)} = w^{(i)} \cdot  a^{(i-1)} + b^{(i)}, \\
\end{split}
\end{equation}
where $i$ is the layer number, $\sigma$ is the activation function, and $w$ and $b$ are the weight and bias of that layer.

Now, as has been seen in~\autoref{fig:phase4}, what is coming back to the client in the backpropagation phase is $\nabla_{Aut(smsh_i)}\mathcal{L}$. However, during a routine training condition, what the client expects to receive is $\nabla_{a^{(i)}}\mathcal{L}$. Due to~\autoref{fig:phase3} and after autoencoder is trained, the attacker aims to see:
\begin{equation}
    Aut(smsh_i) = a'^{(i)} \Rightarrow \nabla_{Aut(smsh_i)}\mathcal{L} = \nabla_{a'^{(i)}}\mathcal{L},
\end{equation}
where $a'^{(i)}$ is considered as an output of an assumed poisoned client (which we expect the output of the autoencoder to be). We also assume that initialized weights of both poisoned and non-poisoned clients are the same (again, for relaxing the assumptions). When injecting a backdoor in a client, one gradient descent step looks like this:
\begin{equation}
        w^{(i)}_{t+1} = w^{(i)}_{t} - \alpha\nabla_{w^{(i)}}\mathcal{L},
\end{equation}
and the attacker aims for \begin{math} w^{(i)}_{t+1} = w'^{(i)}_{t+1} \end{math}, thus:
\begin{equation}
\begin{split}
    w^{(i)}_{t} - \alpha\nabla_{w^{(i)}}\mathcal{L} =  w'^{(i)}_{t} - \alpha\nabla_{w'^{(i)}}\mathcal{L}, \ \Rightarrow \\
    \nabla_{w^{(i)}}\mathcal{L} =  \nabla_{w'^{(i)}}\mathcal{L}, \ \Rightarrow \\
    \nabla_{a'^{(i)}}\mathcal{L} \cdot \frac{\partial a^{(i)}}{\partial z^{(i)}} \cdot \frac{\partial z^{(i)}}{\partial w^{(i)}} = \nabla_{a'^{(i)}}\mathcal{L} \cdot \frac{\partial a'^{(i)}}{\partial z'^{(i)}} \cdot \frac{\partial z'^{(i)}}{\partial w'^{(i)}}, \ \Rightarrow \\
    \nabla_{a'^{(i)}}\mathcal{L} \cdot a^{(i-1)} = \nabla_{a'^{(i)}}\mathcal{L} \cdot a'^{(i-1)}.
\end{split}
\end{equation}

The expressions on the left and right could not be equal together unless we multiply the $\nabla_{a'^{(i)}}\mathcal{L}$ by $\frac{a'^{(i-1)}}{a^{(i-1)}}$. If we go down the layers, the same situation holds, and we have to multiply the flowing gradient to $\frac{a'^{(i-2)}}{a^{(i-2)}}$ and so on back down to input, and since we do not have access to input, the attack success seems unlikely. 

A recent study~\cite{goldwasser2022planting} has theoretically proven that backdoor attacks can be injected into a model using a digital signature scheme. Their study theoretically shows that the possibility of backdooring a supervised learning model is inevitable, and it is computationally infeasible for detection mechanisms that try to defend against such backdoors. 
These novel findings do not contradict our results because the backdooring in~\cite{goldwasser2022planting} is proposed for attacking the whole model, not after the cut layer like ours.
We leave it as a future work to investigate whether it is theoretically possible to plant a backdoor in the upper part of a network without access to training data.   
%-------------------------------------------------------------------------------
\section{Conclusion and Future Work}
\label{sec:conclusion}
%-------------------------------------------------------------------------------

This paper examined the effectiveness of the server-side backdoor attacks on split learning with two methods: a surrogate client and an autoencoder. Experiments were conducted using three model architectures and three image datasets. The results show that split learning is highly resistant to these poisoning attacks. In the SL type of architecture, we believe it is very difficult to inject a backdoor merely by using a server without manipulating or having access to parts of the client side. Although the attack is not sufficiently successful, we conjecture that by relaxing some assumptions (e.g., the attacker gains control over a simple client) in the threat model, there is room for crafting backdoors that could impact the SL structure. The architecture of the SL model has not yet been studied thoroughly, and its susceptibility to poisoning attacks (especially backdoor attacks) needs to be researched more comprehensively.

% \section*{Acknowledgment}

% The preferred spelling of the word ``acknowledgment'' in America is without 
% an ``e'' after the ``g''. Avoid the stilted expression ``one of us (R. B. 
% G.) thanks $\ldots$''. Instead, try ``R. B. G. thanks$\ldots$''. Put sponsor 
% acknowledgments in the unnumbered footnote on the first page.

\bibliographystyle{IEEEtran}
\bibliography{IEEEabrv,bibliography}

\appendices

\section{Additional Experimental Setup Information}
\label{sec:appendix1}
%%%%%%%%%%%%%%%%%%%%%%%%%%%%%%%%%%%%%%%%%%%%%%%%%%%%%%%%%%%%%%
\subsection{Datasets}

We have evaluated our proposed attacks against three publicly available datasets in the image domain:
\begin{itemize}
    \item MNIST: consists of 60000 training and 10000 test images. Each image is a 28$\times$28 gray-scale image of a hand-written digit. It has 10 classes of digits, each having 7000 samples overall (6000 + 1000).
    \item FMNIST: consists of 60000 training and 10000 test images. Each image is a 28$\times$28 gray-scale image of a piece of clothing. It has 10 classes, each having 7000 samples overall (6000 + 1000).
    \item CIFAR10: consists of 50000 training and 10000 test images. Each image is a 32$\times$32 RGB image. It has 10 classes, each having 6000 samples overall (5000 + 1000).
\end{itemize}

\autoref{tab:dataset-split} provides the division of the used datasets for our experiments. 
% \autoref{tab:dataset-split} demonstrate the splitting of the datasets for our experiment. 
The entire validation and test sets are used for the global model performance during training and test time. The test set is used for both evaluating the ASR and CDA. The training set (which is $X\mbox{-}priv_{mlcs} + \bigcup_{i=1}^{n} X\mbox{-}priv_i$) should be divided between all participants, so we split the training set in an i.i.d manner between all participants (which is \#innocent\_clients + 1). When the local dataset for the attacker is delegated, it should be poisoned for using it during an attack. In the surrogate client attack scenario, the attacker poisons 50\% of the local dataset. In the autoencoder injection scenario, in the first step of training, each of the two clients uses the local dataset: one clean and the other 100\% poisoned. In the final phase, which injects the backdoor using an autoencoder, 50\% of incoming smashed data is randomly selected and fed to the autoencoder. The remaining 50\% is just straightly sent to the server.

% \usepackage{multirow}
% \usepackage{multirow}
\begin{table}
\caption{Details of splitting datasets. Training sets are divided by (\# of clients + 1). 1 is the attacker's share from the dataset.}
\label{tab:dataset-split}
\begin{tabular}{|c|c|ccc|}
\hline
Dataset                  & \# of Clients & \multicolumn{3}{c|}{\# of samples}                                                                \\ \hline
                         &               & \multicolumn{1}{c|}{Train} & \multicolumn{1}{c|}{Validation}             & Test                   \\ \hline
\multirow{4}{*}{CIFAR10} & 1             & \multicolumn{1}{c|}{25000} & \multicolumn{1}{c|}{\multirow{4}{*}{5000}}  & \multirow{4}{*}{5000}  \\ \cline{2-3}
                         & 3             & \multicolumn{1}{c|}{12500} & \multicolumn{1}{c|}{}                       &                        \\ \cline{2-3}
                         & 5             & \multicolumn{1}{c|}{8333}  & \multicolumn{1}{c|}{}                       &                        \\ \cline{2-3}
                         & 7             & \multicolumn{1}{c|}{6250}  & \multicolumn{1}{c|}{}                       &                        \\ \hline
\multirow{4}{*}{FMNIST}  & 1             & \multicolumn{1}{c|}{25000} & \multicolumn{1}{c|}{\multirow{4}{*}{10000}} & \multirow{4}{*}{10000} \\ \cline{2-3}
                         & 3             & \multicolumn{1}{c|}{12500} & \multicolumn{1}{c|}{}                       &                        \\ \cline{2-3}
                         & 5             & \multicolumn{1}{c|}{8333}  & \multicolumn{1}{c|}{}                       &                        \\ \cline{2-3}
                         & 7             & \multicolumn{1}{c|}{6250}  & \multicolumn{1}{c|}{}                       &                        \\ \hline
\multirow{4}{*}{MNIST}   & 1             & \multicolumn{1}{c|}{25000} & \multicolumn{1}{c|}{\multirow{4}{*}{10000}} & \multirow{4}{*}{10000} \\ \cline{2-3}
                         & 3             & \multicolumn{1}{c|}{12500} & \multicolumn{1}{c|}{}                       &                        \\ \cline{2-3}
                         & 5             & \multicolumn{1}{c|}{8333}  & \multicolumn{1}{c|}{}                       &                        \\ \cline{2-3}
                         & 7             & \multicolumn{1}{c|}{6250}  & \multicolumn{1}{c|}{}                       &                        \\ \hline
\end{tabular}
\end{table}

\subsection{Model Architectures}

We use three model architectures for performing our attacks: LeNet~\cite{lecun1998gradient}, StripNet~\cite{strip}, and ResNet9~\cite{he2016deep}. We divided and split these networks for our experiments, as shown in~\autoref{tab:model-split}. The split number indicates up to which level the model remains in the client part, and the rest goes to the server.

\begin{table*}
\centering
\caption{Architectures of used models and how they are split. Each split-layer indicates up to which section should be in the client, and the rest of the model in lower rows should go to the server part.}
\label{tab:model-split}
\begin{tabular}{|c|ccc|}
\hline
Split & \multicolumn{3}{c|}{Model}                                                                                                                                                                                                                                                                                                                                                                                                                                                                                                                                                                                                                                    \\ \hline
      & \multicolumn{1}{c|}{Lenet}                                                                                                                                                                  & \multicolumn{1}{c|}{StripNet}                                                                                                                                                                                                                                                                                                                                          & ResNet9                                                                                \\ \hline
1     & \multicolumn{1}{c|}{\begin{tabular}[c]{@{}c@{}}Conv2d(num\_channels, 6, 5),\\ ReLU()\end{tabular}}                                                                                          & \multicolumn{1}{c|}{\begin{tabular}[c]{@{}c@{}}Conv2d(num\_channels, 32, (3, 3), 1)\\ ELU()\end{tabular}}                                                                                                                                                                                                                                                              & Conv\_block(num\_channels, 64, 3, 1)                                                   \\ \hline
2     & \multicolumn{1}{c|}{MaxPool2d(2, 2)}                                                                                                                                                        & \multicolumn{1}{c|}{BatchNorm2d(32)}                                                                                                                                                                                                                                                                                                                                   & Conv\_block(64, 128, 3, 1, pool=true)                                                  \\ \hline
3     & \multicolumn{1}{c|}{\begin{tabular}[c]{@{}c@{}}Conv2d(6, 16, 5)\\ ReLU()\end{tabular}}                                                                                                      & \multicolumn{1}{c|}{\begin{tabular}[c]{@{}c@{}}Conv2d(32, 32, (3, 3), 1)\\ ELU()\end{tabular}}                                                                                                                                                                                                                                                                         & Small\_res\_block(128)                                                                 \\ \hline
Rest  & \multicolumn{1}{c|}{\begin{tabular}[c]{@{}c@{}}MaxPool2d(2, 2)\\ Flatten()\\ Linear(linear\_input\_shape, 120)\\ ReLU()\\ Linear(120, 84)\\ ReLU()\\ Linear(84, num\_classes)\end{tabular}} & \multicolumn{1}{c|}{\begin{tabular}[c]{@{}c@{}}BatchNorm2d(32)\\ MaxPool2d((2, 2))\\ Dropout2d(p=0.2)\\ Conv\_block(32, 64, (3,3), 1)\\ Conv\_block(64, 64, (3,3), 1)\\ MaxPool2d((2, 2))\\ Dropout2d(p=0.3)\\ Conv\_block(64, 128, (3,3), 1)\\ Conv\_block(128, 128, (3,3), 1)\\ MaxPool2d((2, 2))\\ Dropout2d(p=0.4)\\Linear(in\_features, num\_classes)\end{tabular}} & \begin{tabular}[c]{@{}c@{}}MaxPool2d(4 , 4 )\\ Linear(2048, num\_classes)\end{tabular} \\ \hline
\end{tabular}
\end{table*}

\subsubsection{Autoencoder Model Architecture}

Next, we provide the structure of our used autoencoder written in PyTorch.

\begin{adjustbox}{max width=1.0\linewidth}
\begin{lstlisting}[language=Python]

Autoencoder(
  (encoder): Encoder(
    (net): Sequential(
      (0): Conv2d(64, 32, kernel_size=(3, 3), stride=(2, 2), padding=(1, 1))
      (1): GELU(approximate='none')
      (2): Conv2d(32, 32, kernel_size=(3, 3), stride=(1, 1), padding=(1, 1))
      (3): GELU(approximate='none')
      (4): Conv2d(32, 64, kernel_size=(3, 3), stride=(2, 2), padding=(1, 1))
      (5): GELU(approximate='none')
      (6): Conv2d(64, 64, kernel_size=(3, 3), stride=(1, 1), padding=(1, 1))
      (7): GELU(approximate='none')
      (8): Conv2d(64, 64, kernel_size=(3, 3), stride=(2, 2), padding=(1, 1))
      (9): GELU(approximate='none')
      (10): Flatten(start_dim=1, end_dim=-1)
      (11): Linear(in_features=1024, out_features=1024, bias=True)
    )
  )
  (decoder): Decoder(
    (linear): Sequential(
      (0): Linear(in_features=1024, out_features=1024, bias=True)
      (1): GELU(approximate='none')
    )
    (net): Sequential(
      (0): ConvTranspose2d(64, 64, kernel_size=(3, 3), stride=(2, 2),
      padding=(1, 1), output_padding=(1, 1))
      (1): GELU(approximate='none')
      (2): Conv2d(64, 64, kernel_size=(3, 3), stride=(1, 1), padding=(1, 1))
      (3): GELU(approximate='none')
      (4): ConvTranspose2d(64, 32, kernel_size=(2, 2), stride=(2, 2),
      padding=(1, 1))
      (5): GELU(approximate='none')
      (6): Conv2d(32, 32, kernel_size=(3, 3), stride=(1, 1), padding=(1, 1))
      (7): GELU(approximate='none')
      (8): ConvTranspose2d(32, 64, kernel_size=(3, 3), stride=(2, 2),
      padding=(1, 1), output_padding=(1, 1))
      (9): Tanh()
    )
  )
)

\end{lstlisting}
\label{code:autoencoder}
\end{adjustbox}

\section{Baseline Accuracy Results for Trained Models}
\label{sec:appendix3}

\begin{figure}
    \centering
    \includegraphics[width=\linewidth]{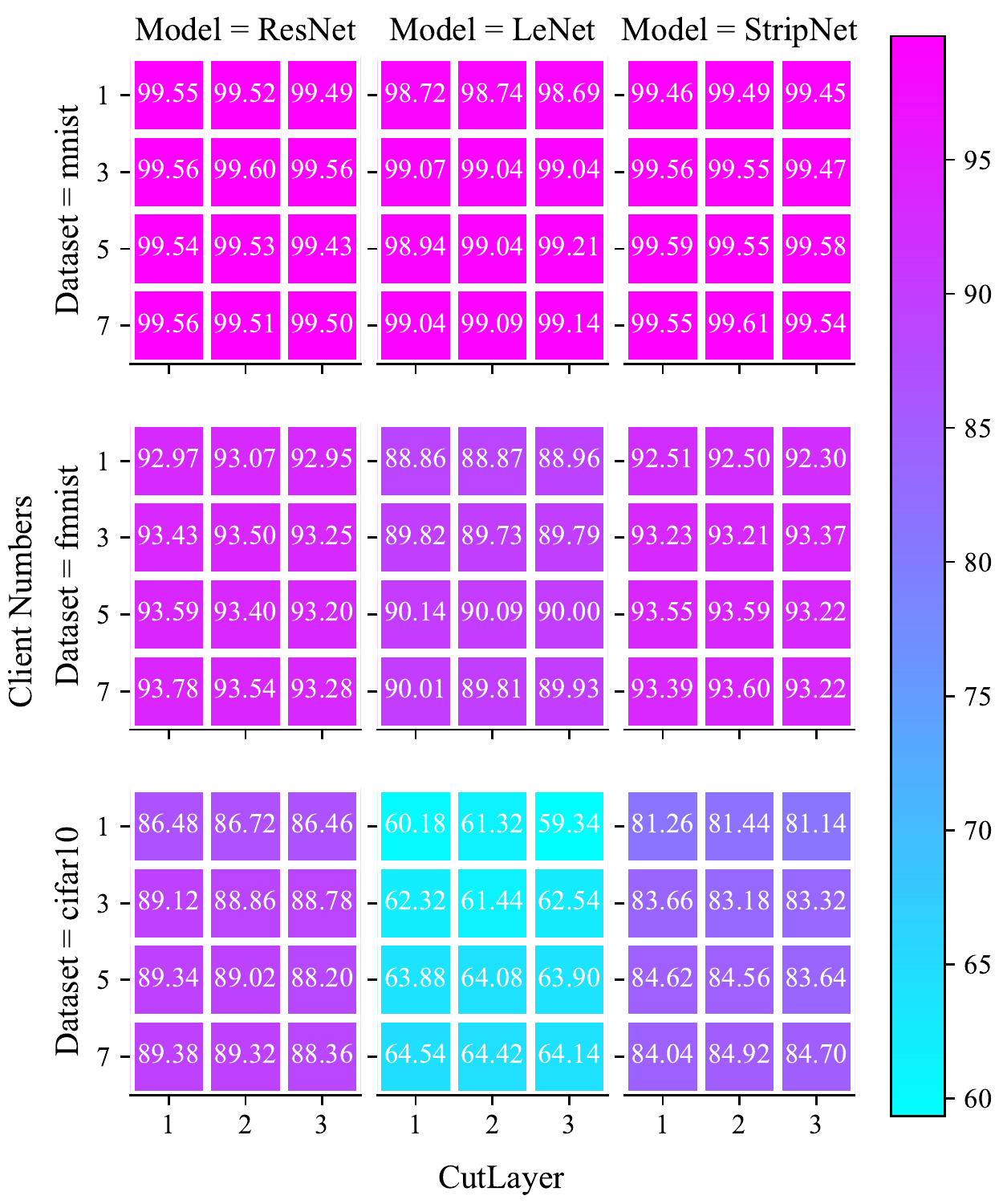}
    \caption{Baseline accuracy for trained model on clean datasets.}
    \label{fig:baseline_clean_acc}
\end{figure}

\autoref{fig:baseline_clean_acc} demonstrates the baseline accuracies for the models we trained in various settings for our experiments. As shown, with CIFAR10, the models achieve the lowest accuracy, while the best performances are on MNIST.

% \section{Split Learning Training Procedure}
% \label{sec:appendix2}

% \subsection{HFL training procedure}
% \label{sec:hfl_training_prcss}

% Several implementations of Federated Learning have been proposed. In the most clear-cut setup, there is a central server $S$, which is responsible for collecting the parameters and updating the global model. Considering the hospital example mentioned earlier, the training protocol is as follows:
% \begin{enumerate}
%     \item In the initial setup phase and after choosing a learning task by all participants, the central server initiates and shares a global model $M$ and its parameters (usually pre-trained) among remote clients.
%     \item Then, at each training step $t$:
%     \begin{enumerate}
%         \item Each remote client $c_i$ downloads the updated global model $M^t$ and, for multiple iterations, trains its local model $m_i$ using its own private dataset $X\mbox{-}priv_i$. When the training is finished, the remote client sends $\Delta\theta_i^t=\theta_i^{t+1} - \theta_M^t$ back to the server.
%         \item After receiving parameter updates ($\Delta\Theta^t=\{\Delta\theta_1^t, ..., \Delta\theta_n^t\}$) from remote local clients, the server updates the global model using its chosen aggregation algorithm ($M^{t+1}=AGG(\Delta\Theta^t)$) and shares the new model.
        
%     \end{enumerate}
% \end{enumerate}

\section{SL Training Procedure}
\label{sec:appendix2}

\begin{figure}[t!]
  \centering
  %\begin{subfigure}[b]{0.75\linewidth}
    \includegraphics[width=\linewidth]{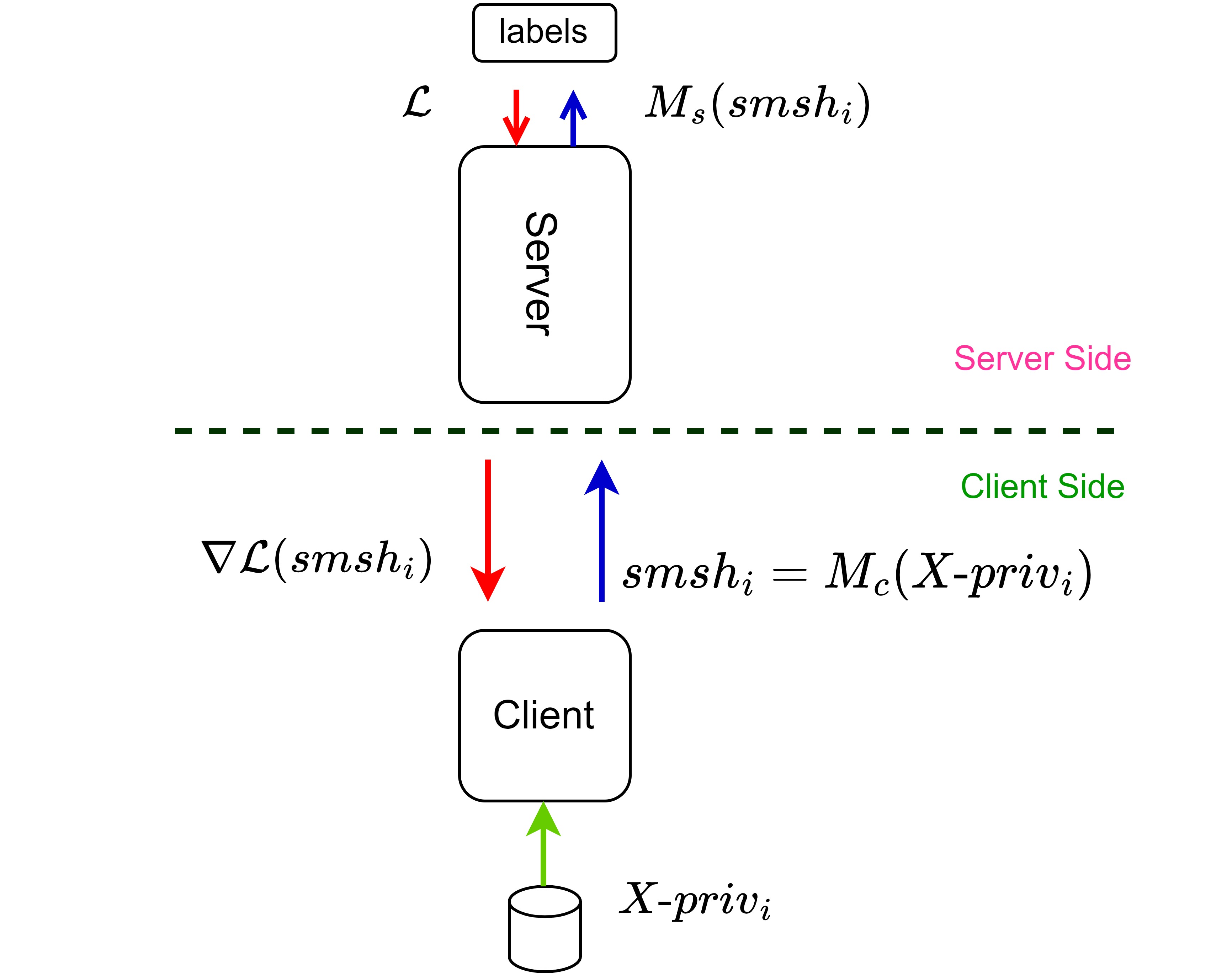}
    % \caption{Vanilla SL}
    % \label{fig:Vanilla_SL_training}
  % \end{subfigure}\\
  % \begin{subfigure}[b]{0.75\linewidth}
  %   \includegraphics[width=\linewidth]{img/SL_Boomerang.jpg}
  %   \caption{Boomerang SL}
  %   \label{fig:Boom_SL_training}
  % \end{subfigure}
  \caption{Training procedure for the vanilla design of SL.}
  \label{fig:SL_training}
\end{figure}

\autoref{fig:SL_training} demonstrates a typical procedure of training the SL model for vanilla architectures.

\section{Why Does the Attack Succeed on MNIST?}
\label{sec:appendix4}

\begin{figure}
    \centering
    \includegraphics[width=0.8\linewidth]{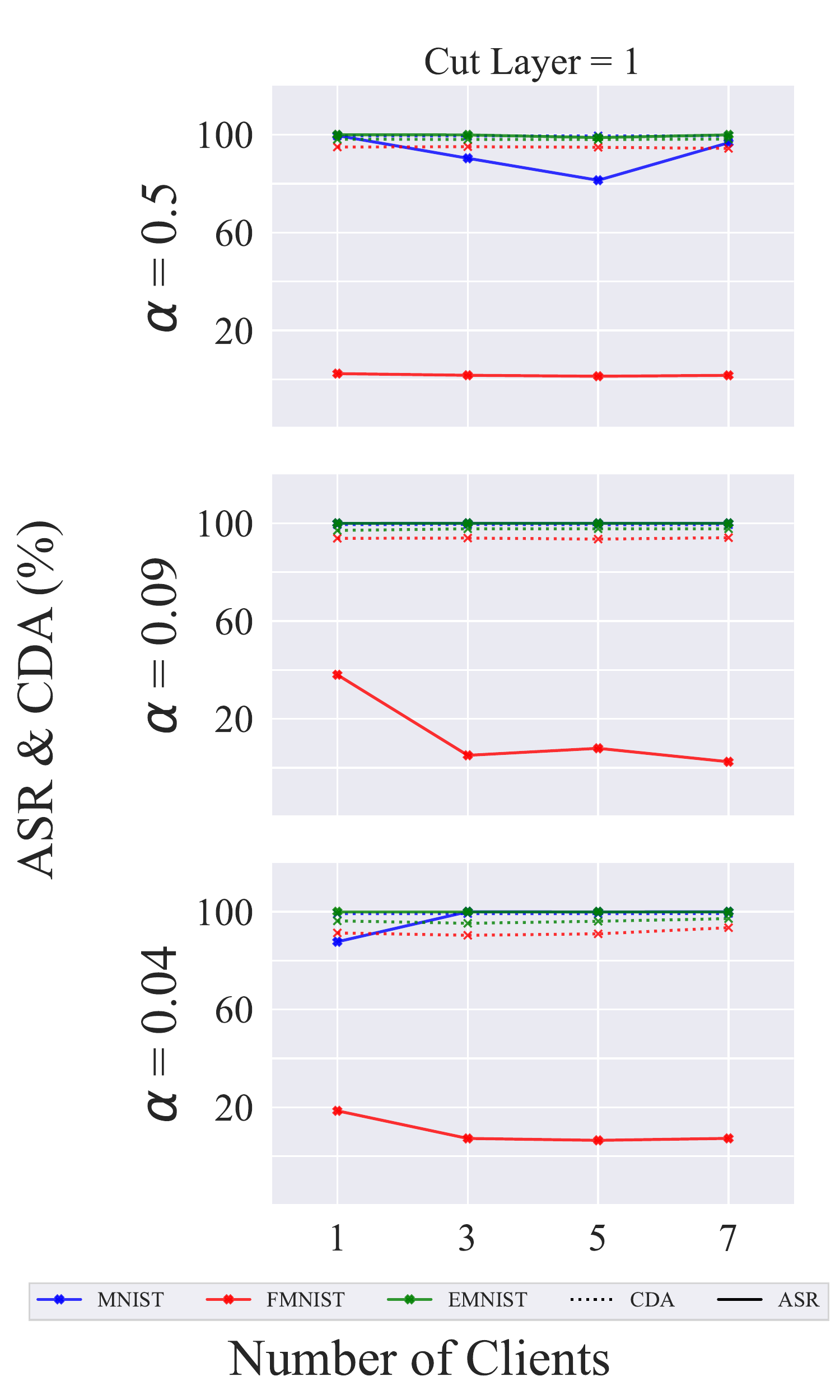}
    \caption{ASR and CDA for the EMNIST (A-J), MNIST (10000 training samples), and FMNIST (only first 5 classes).}
    \label{fig:asr_sheperd}
\end{figure}

First, we emphasize that we run the experiments with strong assumptions (e.g., large $8\times8$ trigger pattern). For instance, in our initial experiments using a $4\times4$ trigger pattern, even the MNIST dataset failed to achieve a promising ASR (with an ASR of $30\%$ as our best result).\\
For the surrogate client attack, we got promising results for the MNIST dataset but not for FMNIST or CIFAR10. As mentioned in \autoref{sec:discussion}, for a backdoor trigger to be effectively implanted within the client component, it is imperative that the client is exposed to poisoned input. This ensures that during the optimization process, the weight and bias updates through gradient descent algorithms consider gradients computed based on the poison loss rather than the loss from clean data. If this condition is unattainable, the attacker would require direct access to the lower-level parameters to manipulate them. \\
Regardless of the method employed, upon the completion of training, the client component must function in a manner that yields outputs analogous to a poisoned client (i.e., $M_C(x_{poisoned}) \simeq M_{C^{'}}(x_{poisoned})$ during test time where $C$ and $C^{'}$ are normal and poisoned clients after training, respectively). The probability of this happening by accident at the end of training is negligible. Given that the attacker cannot access either the input or the lower-level network (client), we do not anticipate the attack to be successful, even with the MNIST dataset. \\
We already know that for the attack to be effective during testing, specific patterns in the smashed data values must exist to activate the infected neurons that lie dormant within the remaining network (i.e., the backdoor previously embedded in the upper server component). By putting just a few layers on the client side and leaving the rest on the server, we assume that the forward pass for MNIST input is simple enough to preserve the trigger inside the smashed data even if the client part has not seen and memorized the poisoned input during the training phase (this is particularly evident in the case of convolutional layers, where the underlying structural organization of the convolutional kernels enables the preservation of input patterns within the resulting feature maps to a considerable extent). Consequently, since the majority of the network resides on the server side, which is already poisoned, activation can occur, enabling the attack to succeed.\\
To address the concerns raised, we conducted additional experiments to test our hypothesis. We repeated the attack with three datasets to explore the factors that could affect ASR: the EMNIST dataset with the same number of classes as MNIST (we chose the letters A to J). 
We aimed to see whether we get the same ASR and behavior from the network since MNIST and EMNIST are equivalent in terms of the complexity and nature of their samples. We also repeated the experiment with FMNIST with the first five classes to see if the number of classes makes it difficult for the backdoor to be injected into the network for more complex datasets like FMNIST and CIFAR10. 
We used the same MNIST dataset for the third experiment but reduced the training size to 10000 samples to see if fewer samples lead to low ASRs. We chose $cutlayer=1$ and StripNet as our best settings for which we got the best ASR. \autoref{fig:asr_sheperd} demonstrates the results for mentioned experiments. EMNIST got almost the same results as MNIST, as this dataset has the same features as MNIST (same dimension, channels, range of grayscale values, etc.). For FMNIST, the ASR values are not promising again, which shows the number of classes does not determine whether the backdoor pattern could be injected successfully. Reduction in the number of training samples does not affect the ASR, as seen from the MNIST ASR values. \\
Based on our assumptions and experiments, we conjecture that the nature of the MNIST grayscale dataset is not reliable enough for research experiments similar to ours. It is so simple that probably any little activation from the client can activate the trigger planted inside the server side. \\ 
Whether designing more powerful and dataset-specific triggers could do the same effect on other, more complex dataset is an interesting research question that should be considered in future works.

\section{Does the Backdoor Remain during Transfer Learning?}
\label{sec:appendix5}
To further explore whether our successful attack on MNIST can be preserved in the model and transferred via another task, we run our experiments in the transfer learning scheme. For this, we chose the model with the best ASR: StripNet model, which is backdoored by training on the MNIST dataset, with 3 clients, $alpha=0.04$, and $cutlayer=1$. The attacker achieves $100\%$ ASR in the mentioned setting. Then, we assemble back the server and client parts as a whole model and freeze all layers except the final fully connected one. We train the model again on 3 clean datasets and test them on the corresponding poisoned dataset.
\autoref{tab:transfer_lr} demonstrates the results of our transfer learning task. The backdoor fails to transfer for a new task even inside the MNIST dataset. We can conclude that planted backdoor inside of the model is not robust enough to remain effective after a transfer learning task, and the SL structure remains safe against this attack concerning transferability.

\begin{table}
\caption{ASR and CDA results for the transfer learning task. We used a backdoored StripNet model trained on the MNIST dataset with $100\%$ ASR.}
\label{tab:transfer_lr}
\begin{tabularx}{0.8\linewidth}{X|XX}

Dataset & CDA   & ASR   \\ \hline
MNIST   & 99.45 & 57.41 \\ \hline
EMNIST  & 95.77 & 0.25  \\ \hline
FMNIST  & 87.02 & 3.2   \\ \hline
\end{tabularx}
\end{table}

\end{document}